\documentclass[aps,prl,preprint,preprintnumbers,showpacs,superscriptaddress,showpacs]{revtex4-1}

\usepackage{graphicx}
\usepackage{times} 
\usepackage[normalem]{ulem}
\usepackage[dvips]{color}
\usepackage{color}
\usepackage{amsmath, amsthm, amssymb}

\begin{document}

\title{
Spin precession mapping at ferromagnetic resonance via nuclear resonant scattering}

\author{Lars Bocklage}
\email{lars.bocklage@desy.de}
\affiliation{Deutsches Elektronen-Synchrotron, Notkestra{\ss}e 85, 22607 Hamburg, Germany}
\affiliation{The Hamburg Centre for Ultrafast Imaging, Luruper Chaussee 149, 22761 Hamburg, Germany}

\author{Christian Swoboda}
\affiliation{Institut f\"ur Angewandte Physik und Zentrum f\"ur
Mikrostrukturforschung, Universit\"at  Hamburg, Jungiusstrasse 11, 20355
Hamburg}
\affiliation{The Hamburg Centre for Ultrafast Imaging, Luruper Chaussee 149, 22761 Hamburg, Germany}

\author{Kai Schlage}
\affiliation{Deutsches Elektronen-Synchrotron, Notkestra{\ss}e 85, 22607 Hamburg, Germany}

\author{Hans-Christian Wille}
\affiliation{Deutsches Elektronen-Synchrotron, Notkestra{\ss}e 85, 22607 Hamburg, Germany}

\author{Liudmila Dzemiantsova}
\affiliation{Deutsches Elektronen-Synchrotron, Notkestra{\ss}e 85, 22607 Hamburg, Germany}
\affiliation{The Hamburg Centre for Ultrafast Imaging, Luruper Chaussee 149, 22761 Hamburg, Germany}

\author{Sa\v{s}a Bajt}
\affiliation{Deutsches Elektronen-Synchrotron, Notkestra{\ss}e 85, 22607 Hamburg, Germany}

\author{Guido Meier}
\affiliation{Max-Planck Institute for the Structure and Dynamics of Matter, Luruper Chaussee 149, 22761 Hamburg, Germany}
\affiliation{Institut f\"ur Angewandte Physik und Zentrum f\"ur
Mikrostrukturforschung, Universit\"at  Hamburg, Jungiusstrasse 11, 20355
Hamburg}
\affiliation{The Hamburg Centre for Ultrafast Imaging, Luruper Chaussee 149, 22761 Hamburg, Germany}

\author{Ralf R\"ohlsberger}
\affiliation{Deutsches Elektronen-Synchrotron, Notkestra{\ss}e 85, 22607 Hamburg, Germany}
\affiliation{The Hamburg Centre for Ultrafast Imaging, Luruper Chaussee 149, 22761 Hamburg, Germany}

\date{\today}

\begin{abstract}
We probe the spin dynamics in a thin magnetic film at ferromagnetic resonance by nuclear resonant scattering of synchrotron radiation at the 14.4 keV resonance of $^{57}$Fe. The precession of the magnetization leads to an apparent reduction of the magnetic hyperfine field acting at the $^{57}$Fe nuclei. The spin dynamics is described in a stochastic relaxation model adapted to the ferromagnetic resonance theory by Smit and Beljers to model the decay of the excited nuclear state. From the fits of the measured data the shape of the precession cone of the spins is determined. Our results open a new perspective to determine magnetization dynamics in layered structures with very high depth resolution by employing ultrathin isotopic probe layers.

\end{abstract}


\maketitle

Spin waves, collective excitations of the magnetization, are key features of magnetic materials as they determine switching times and energy losses during magnetization reversal. They play a crucial role for new concepts of logical operations, spin-torque oscillators, or signal processing, and constitute the basis of the emerging field of magnonics~\cite{KruglyakJOPD2010}. Spin waves are often probed by inelastic scattering techniques like Raman scattering, Brillouin light scattering, or inelastic neutron scattering that rely on analysis of the energy transfer to the scattered particles. We show that resonant spin dynamics can be probed by coherent elastic scattering, namely nuclear resonant scattering of synchrotron radiation (NRS)~\cite{GerdauPRL1985,RoehlsbergerSpringer2005}. Since NRS is an x-ray technique, it opens the possibility to combine it with diffraction methods to obtain very high spatial resolution~\cite{SchlageNJP2012} down to atomic length scales~\cite{GerdauPRL1985}. Moreover, as NRS probes the nuclear decay of an excited M\"ossbauer isotope it allows to employ ultrathin isotopic probe layers to achieve sub-nm depth resolution~\cite{RoehlsbergerPRL2002}.

The application of conventional M\"ossbauer spectroscopy as well as NRS has given indirect access to thermally excited spin waves in ferromagnets via the temperature dependence of the magnetic hyperfine field that exhibits the same temperature dependence as the magnetization in these materials~\cite{EibschuetzPRB1973,ChienPRB1977,SenzNJP2003}. In contrast to a broad thermally excited spin wave spectrum we induce a single 
coherent mode by resonant excitation with radio frequency (rf) magnetic fields. This is one approach typically used in magnonic devices to obtain the desired functionality~\cite{DemokritovNAT2006,KostylevAPL2005}.

The impact of rf magnetic fields on conventional M\"ossbauer spectroscopy has so far been studied at MHz frequencies that are well below ferromagnetic resonances. It was shown that low-frequency spin waves in paramagnetic media can induce magetic phase modulations of the nuclear states~\cite{SinorPRL1989}. A collapse of the magnetic hyperfine field arises from a fast periodic switching of the magnetization~\cite{PfeifferJAP1971,KopcewiczSC1991}. The sideband effect originates from acoustic vibrations induced by magnetostriction~\cite{HeimanPRL1968,KopcewiczSC1991}. Also Rabi oscillations have been observed where the rf field directly couples to the nuclear transition~\cite{TittonenPRL1992}. Ferromagnetic or spin wave resonances at GHz frequencies have not been investigated with M\"ossbauer spectroscopy or NRS so far. This is the case to be studied here.

In this letter we show that NRS measurements enable to extract the trajectory of the spins during coherent precession at ferromagnetic resonance. The opening angle of the precession is an essential parameter for spintronics like, e.g., in spin pumping~\cite{TserkovnyakPMP2005,KuhlmannPRB2013} or in the spin dynamo~\cite{GuiPRL2007}. So far the determination of the opening angle from experiments appeared to be challenging. Values for opening angles averaged over time and space have been determined by anisotropic magnetoresistance measurements for assumed circular trajectories~\cite{CostacheAPL06b,KuhlmannPRB2012}, an assumption that is justified only in a few special cases like in a spherical particles without crystalline anisotropy.

We performed NRS measurements on ferromagnetic thin films excited at ferromagnetic resonance. The influence of spin dynamics on the NRS signal is analyzed in a stochastic relaxation model~\cite{BlumePR1968b,ClauserPRB1971} adapted to the ferromagnetic resonance theory of Smit and Beljers~\cite{SmitPRR1955}. With this method the exact shape of the precession orbit is determined. This capability arises from the high sensitivity of NRS to the magnetization direction.  

NRS under grazing incidence~\cite{RoehlsbergerSpringer2005} is performed at the Dynamics beamline~P01~\cite{WilleJPCS2010} of PETRA~III at DESY (Hamburg, Germany) in 40~bunch mode with a bunch separation of 192~ns and a bunch duration of about 50~ps. The energy is tuned to the nuclear transition energy of 14.4125 keV of the M\"ossbauer isotope $^{57}$Fe that has a natural lifetime of 141.11~ns. The bandwidth of the synchrotron radiation is reduced to 1~meV by a high resolution monochromator. A Kirkpatrick-Baez multilayer mirror system focuses the beam to a spot size of about 10~$\times$~10~$\mu$m$^2$. The synchrotron pulse excites all six allowed nuclear transitions simultaneously. The frequency differences manifest as quantum beats in the temporal evolution of the nuclear decay. The recorded NRS time spectra are fingerprints for the magnetic spin structure of the sample.

\begin{figure}
\includegraphics{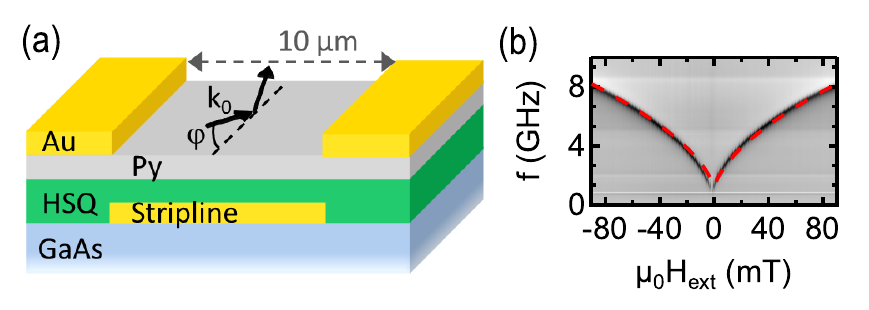}
\caption{\label{Fig1} (color online) (a) Schematic layer system of the sample. The angle of incidence of the photon wave vector $k_0$ is $\varphi$. (b) Absorption spectrum of the permalloy film depending on external field and excitation frequency. Black indicates high absorption. Dashed red line is the fit to the Kittel formula.}
\end{figure}

Samples are prepared on GaAs substrates. All layer geometries are defined by electron-beam lithography and lift-off processing. Electrical striplines are prepared by thermal evaporation of 7~nm Cr, 118~nm Ag, and 20~nm Au. Hydrogen silsesquioxane (HSQ) with a thickness of 140~nm is used to electrically insulate the strip line from the ferromagnetic film and to provide a smooth surface. The 800~$\times$~800~$\mu$m$^2$ film is prepared by sputter deposition of 4~nm Cr, 18~nm Pd, 13~nm isotopically enriched permalloy (Ni$_{81}\, ^{57}$Fe$_{19}$), and 2~nm Pd. A scheme of the layer system is shown in Fig.~\ref{Fig1}(a). The stoichiometry of permalloy is confirmed by energy-dispersive x-ray spectroscopy. The sideband effect is negligible in permalloy due to the low magnetostriction and the high frequencies~\cite{TittonenPRL1992}. Because the ferromagnetic resonance of the film is only effectively excited right above the 10~$\mu$m wide stripline, any NRS signal from non-excited parts of the magnetic film is blocked by a highly x-ray absorbing bilayer of 7~nm Al and 30~nm Au on top of the magnetic film. The sample is illuminated under grazing-incidence at an angle~$\varphi$ of 4.36~mrad, the critical angle for total reflection where the nuclear signal reaches its maximum~\cite{Supplementary}. The 10~$\times$~800~$\mu$m$^2$ part of permalloy film is completely illuminated by the microbeam under grazing incidence.

External fields are applied in the plane of the film. The stripline is connected to a vector network analyzer that serves as a signal source for the high frequency excitation as well as detector for the transmitted signal. Figure~\ref{Fig1}(b) shows an electrical absorption spectrum of the film where high absorption (black) indicates the excitation of the ferromagnetic resonance (Kittel mode) of the permalloy film~\cite{KittelRMP49}, a spin wave mode with zero wave vector that corresponds to a uniform precession of the magnetic moments. A fit with the Kittel formula yields a saturation magnetization of $M_{\text{S}} =$~666~kA/m and a damping parameter of $\alpha = $~0.012.

\begin{figure}
\includegraphics{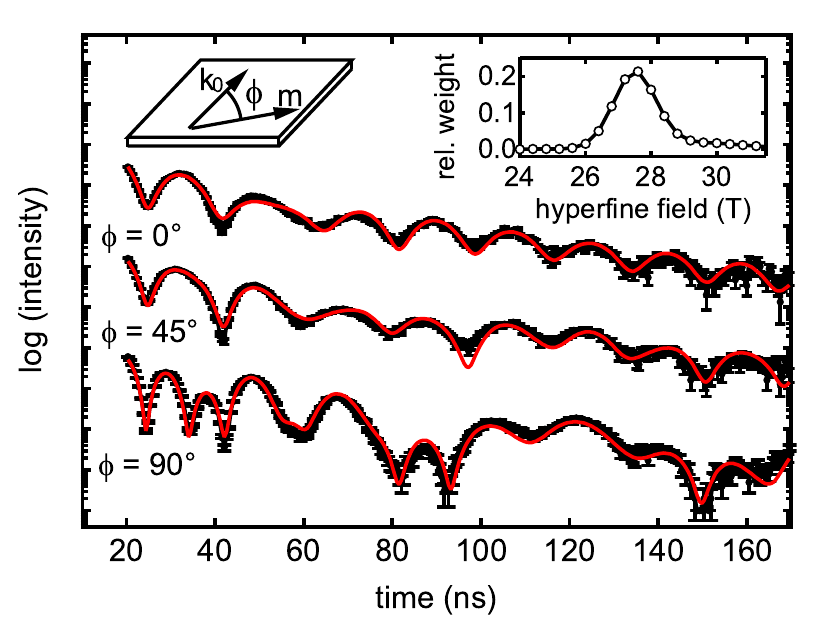}
\caption{\label{Fig2} (color online) NRS time spectra from the permalloy film at different in-plane angles $\phi$ between the photon wave vector $k_0$ and the magnetization $m$. The red lines show fits calculated with the NRS evaluation package CONUSS. Curves are offset for clarity. The deduced hyperfine field distribution of the permalloy film is given in the inset.}
\end{figure}

The precise evaluation of the NRS time spectra requires the exact knowledge of the hyperfine field distribution in the sample. This can be obtained from the evaluation of NRS time spectra without rf magnetic field excitation at different in-plane angles $\phi$ between the incoming photon wave vector $k_0$ and an external field of 70~mT. Because of the low coercivity of the permalloy film of less than 1~mT~\cite{KamionkaPRB2011} the magnetization $m$ and the external field are parallel. The NRS time spectra are shown in Fig.~\ref{Fig2} complemented with fits using the NRS evaluation package CONUSS~\cite{SturhahnHI2000}. From the calculations the hyperfine field distribution of the permalloy film at about 27.6~T is deduced as shown in the inset of Fig.~\ref{Fig2}. Isomer shift and quadrupole splitting are 0.134~mm/s and 0.020~mm/s, respectively, similar to values found in previous studies on permalloy~\cite{PingJMMM1992}.

\begin{figure}
\includegraphics{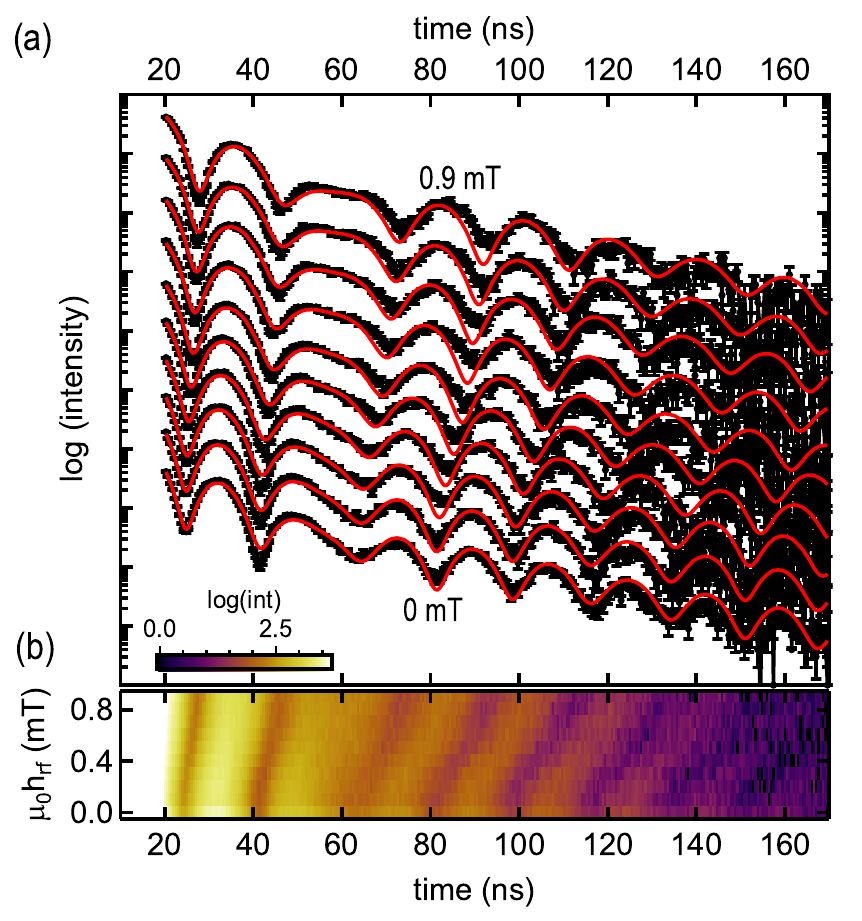}
\caption{\label{Fig3} (color online) (a) Time spectra at various excitation field amplitudes. The lowest time spectrum is not excited. The excitation field $\mu_0h_{\text{rf}}$ increases up to 0.9~mT in steps of 0.1~mT. Curves are offset for clarity. Red lines are fits with CONUSS. 
(b) Logarithmic intensity map of the time spectra shown in (a) with varying excitation field $h_{\text{rf}}$.}
\end{figure}

In the following the Kittel mode is excited at different static fields applied parallel to the incoming beam. Time spectra for a constant external field of 5~mT at different radio-frequency magnetic field amplitudes $h_{\text{rf}}$ are shown in Fig.~\ref{Fig3}(a). The resonance frequency is 1.93~GHz as determined from the electrical absorption measurements. The overall shape of the time spectra alters with increasing excitation field. A shift of the extrema to later times with increasing dynamic field is visualized in Fig.~\ref{Fig3}(b). In addition a slightly faster decay of the time spectra is observed during ferromagnetic resonance. The shift of the extrema indicates that the effective magnitude of the hyperfine field is lower. It results in a reduction of the magnetic hyperfine splitting and a correspondingly larger period in the temporal beat pattern. The effect resembles the temperature dependence of the hyperfine field that originates from thermal excitation of spin waves~\cite{EibschuetzPRB1973,ChienPRB1977,SenzNJP2003}. However, here we excite only one coherent mode which allows to determine dynamic magnetic properties under conditions of ferromagnetic resonances.


The hyperfine interaction energy $E_{\text{hf}} = (m_{\text{e}} g_{\text{e}}-m_{\text{g}} g_{\text{g}})B_{\text{hf}} = \hbar \omega_{\text{hf}}$ determines the time scale for dynamic effects on the NRS signal~\cite{BlumePR1968,LeopoldHI1999}. Here $m_{\text{g}}$ and $m_{\text{e}}$ are the magnetic quantum numbers and $g_{\text{g}}$ and $g_{\text{e}}$ are the g-factors of the ground and excited states, respectively. For hyperfine fields observed in permalloy of 27.6~T~ the frequency $\omega_{\text{hf}}/2\pi$ is in the order of a few ten MHz. For any change of the hyperfine field much faster than $1/\omega_{\text{hf}}$ the nucleus cannot follow the hyperfine field and it experiences an effective hyperfine field resulting from temporal averaging over the relatively long life time of the nuclear excited state. This situation corresponds to the fast relaxation regime. 
As a consequence a reduced effective hyperfine field is observed for a precession of the magnetization around the equilibrium direction. Moreover, the high sensitivity of NRS to the magnetization direction, as shown in Fig.\ref{Fig2}, yields the capability to determine the precession orbit of the magnetization because its dynamic in-plane and out-of-plane components influence the time spectra.

\begin{figure}
\includegraphics{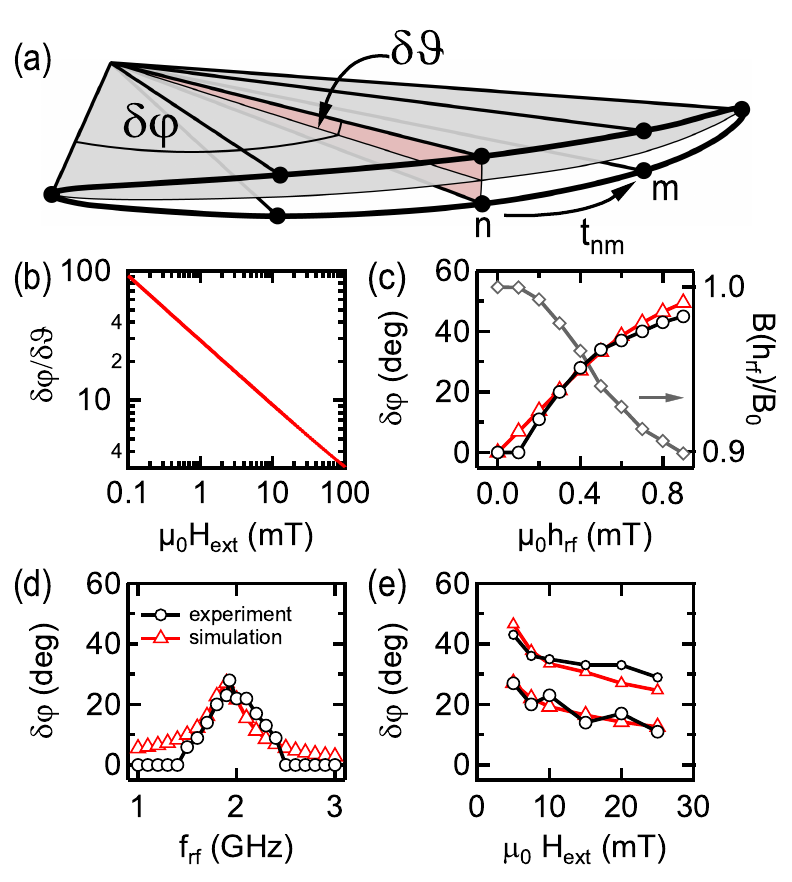}
\caption{\label{Fig4} (color online) (a) Scheme of the discretization of the spin wave precession cone for the stochastic relaxation model. Gray indicates the film plane and red the out-of-plane direction. $\delta\varphi$ and $\delta\theta$ are the dynamic in-plane and out-of-plane angles, respectively. The black dots represent the magnetization directions with transition rates $t_{nm}$ used in the CONUSS calculations of the NRS time spectra.
(b) Ratio of the dynamic angles for the permalloy film.
(c)-(e) Maximum in-plane angle $\delta\varphi$ of the spin precession cone in dependence of the excitation field amplitude at an external field of 5~mT with a resonance frequency of 1.93~GHz (c), of the excitation frequency at a field amplitude of 0.4~mT (d), and of the external field at the resonance frequency with rf magnetic field amplitudes of 0.4~mT and 0.8~mT (smaller symbols)(e). Data deduced with CONUSS from the measurements (circles) and data from micromagnetic simulations (triangles) are shown. In (c) the reduced hyperfine field is given in dependence of the rf magnetic field (rhombi). Lines are guides to the eye.}
\end{figure}

For a quantitative evaluation of the influence of the spin wave on the time spectra calculations within the stochastic relaxation model are performed~\cite{BlumePR1968b,ClauserPRB1971} that is implemented in CONUSS to account for dynamics of the hyperfine field. The stochastic model assumes discrete hyperfine field directions with transition rates $t_{\text{nm}}$ between field directions $n$ and $m$. We model the spin precession by eight points on the precession cone as shown in Fig.~\ref{Fig4}~(a). The transition matrix is chosen to allow transitions between neighboring points in a way that supports only one sense of rotation as realized in the experiment~\cite{Supplementary}. The transition rates equal the inverse period of the spin wave meaning that every transition is performed once per cycle.

The field directions on the precession trajectory have to be modeled for the CONUSS calculations. In the thin film the demagnetization field has to be considered that leads to an elliptical precession. We deduce the precession trajectory of the magnetization from the Smit-Beljers formulation of the ferromagnetic resonance~\cite{SmitPRR1955}. The Smit-Beljers formulation yields a set of coupled differential equations for the dynamic components $\delta\varphi$ and $\delta\vartheta$ of the azimuthal angle $\varphi$ and polar angle $\vartheta$ of the magnetization (see Fig.~\ref{Fig4})~\cite{SkrotskiiFR1966}
\begin{eqnarray}
\label{eq1}
     -\gamma^{-1} M_{\text{S}} \sin(\vartheta_0) \delta\dot{\vartheta} & = & F_{\varphi\vartheta}\delta\vartheta +  F_{\varphi\varphi}\delta\varphi \nonumber\\
  \gamma^{-1} M_{\text{S}} \sin(\vartheta_0) \delta\dot{\varphi} & = & F_{\vartheta\vartheta}\delta\vartheta +  F_{\vartheta\varphi}\delta\varphi
\end{eqnarray}
where $\gamma$ is the gyromagnetic ratio, $\vartheta_0$ is the equilibrium polar angle, and $F$ is the free energy density. The indices of $F$ indicate partial derivatives at equilibrium positions. From Eq.~(\ref{eq1}) we derive the ratio of the dynamic angles at resonance
\begin{equation}
\label{ratio}
\dfrac{\delta\varphi}{\delta\vartheta} = - \dfrac{F_{\varphi\vartheta}+i \sqrt{F_{\varphi\varphi}F_{\vartheta\vartheta}-F_{\varphi\vartheta}^2}}{F_{\varphi\varphi}}\;.
\end{equation}
To get the correct ratio for our evaluation the angular components have to be decoupled ($F_{\varphi\vartheta} = F_{\vartheta\varphi} = 0$) meaning that the magnetization lies in the azimuthal plane. Then, Eq.~(\ref{ratio}) simplifies to $\delta\varphi/\delta\vartheta = -i \sqrt{F_{\vartheta\vartheta}/F_{\varphi\varphi}}$.

For a thin film magnetized in the plane without crystalline anisotropy the free energy density is $F=-\mu_0M_{\text{S}} H_{ext}\sin\vartheta\cos \varphi+\frac{1}{2}\mu_0M_{\text{S}}^2 \cos^2\vartheta$, where $\varphi$ is the in-plane angle between magnetization and external field. At equilibrium the magnetization and external field are parallel and we get
\begin{equation}
	\left|\frac{\delta\varphi}{\delta\vartheta}\right| = \sqrt{\frac{H_{\text{ext}}+M_{\text{S}}}{ H_{\text{ext}}}}.
\end{equation}
This ratio for the film, shown in Fig.~\ref{Fig4}(b), is inserted in the CONUSS calculations for the parametrization of the precession cone. For example, the experimental conditions presented in Fig.~\ref{Fig3} with an external field of 5~mT yield a ratio of 13.0 (see Fig.~\ref{Fig4}(b)). The maximum dynamic in-plane angle $\delta\varphi$ is the only free parameter in the fits of the time spectra under spin wave excitation.

Figure~\ref{Fig4}(c)-(e) show the in-plane opening angle $\delta\varphi$ deduced from the time spectra. Fig.~\ref{Fig4}(c) and (d) display data at an external field of 5~mT where the resonance frequency is 1.93~GHz. For increasing excitation fields the opening angle increases as well, while the gain is lower at higher fields (Fig.~\ref{Fig4}(c)).
With an excitation field of 0.4~mT the opening angles around resonance in Fig.~\ref{Fig4}(d) are obtained. The resonant behavior clearly shows that the hyperfine field does not decrease due to Joule heating or eddy currents. Figure~\ref{Fig4}(e) shows the opening angle at resonance for different static external fields. For increasing fields the opening angle is reduced.

The values of the extracted dynamic angles are compared to micromagnetic simulations. These simulations have been carried out with the program package MicroMagnum~\cite{MicroMagnum} with a film thickness of 12.9~nm using a cell size of 5 x 5 x 4.3 nm$^3$, a saturation magnetization of 666 kA/m, a Gilbert damping of 0.012, and periodic boundary conditions for the x and y direction. The external field has been slightly adjusted in the simulation to fit the experimentally observed resonance frequency. The opening angles calculated from the simulations are also shown in Fig.~\ref{Fig4}(c)-(e). The calculated angles assort well with the values obtained from the NRS spectra. The excellent agreement of the measured and simulated data for the thin film demonstrates how powerful NRS is to sense spin dynamics. Especially for nanoscaled samples, where theoretical modeling might get difficult, NRS could provide distinct advantages in determining dynamic properties.

The measurements indeed allow to distinguish between different assumed trajectories. The simplest model is a circular precession cone. This model fits the data quite reasonably because the average hyperfine field is the projection on the equilibrium direction of the magnetization. The slightly faster decay of the NRS time spectra due to the dynamics is reproduced as well. However, the dynamic in-plane and out-of plane components of the magnetization precession change for the circular cone compared to the elliptical trajectory. This change generates differences in the fit quality and yields poorer fits with least squares up to 9\% higher. 
However, we can simply deduce the reduction of the effective hyperfine field from the projection of the circular cone on the hyperfine field axis. The in this way calculated reduced hyperfine field $B_{hf}(h_{rf})/B_{hf}(0)$ is shown in Fig.~\ref{Fig4}(c). At an excitation field of 0.9~mT a reduction of 10\% is obtained. For small opening angles the hyperfine field reduction becomes less significant. Because the effective hyperfine field scales with the average of the cosine of the dynamic angles the change of the hyperfine field is tiny. For the smallest deduced in-plane angle of 6$^{\circ}$, the out-of-plane angle is 0.46$^{\circ}$ and we get hyperfine field reductions of 0.55~\% and 0.003~\%, respectively. However, these small reductions are sufficient to induce changes of the NRS time spectra compared to the static case.


In summary we have shown that the magnetic hyperfine field is significantly reduced at ferromagnetic resonance. From this reduction one can deduce the cone angle of the spin precession in the thin fim. The measured hyperfine field reduction should be present for all kinds of spin dynamics not only at ferromagnetic resonance. The technique is also applicable to non-zero wave vectors like propagating or confined spin waves due to the nature of the scattering process. The overall good agreement of the micromagnetic simulations and the evaluated data demonstrate the feasibility to study spin dynamics with high accuracy via NRS. The method's isotopic-sensitivity can be employed to study depth profiles and interface effects~\cite{RoehlsbergerPRL2002} related to spin dynamics. Combination of grazing-incidence diffraction and time-resolved NRS will enable three-dimensional mapping of spin waves confined in nanostructures.





We thank U. Merkt and T. Matsuyama for persistent encouragement and fruitful discussions, M. Volkmann and A. Berg for excellent technical support, A. Aquila for software code for KB mirror calculations, A. Rothkirch for help during the data processing, C. Adolff for support during lithography, J. Major for production of a sputter target and D. Schumacher as well as T. Guryeva for sputter deposition. Financial support of the Deutsche Forschungsgemeinschaft via excellence cluster `The Hamburg Centre for Ultrafast Imaging - Structure, Dynamics and Control of Matter on the Atomic Scale', via Sonderforschungs\-be\-reich 668, and via Graduiertenkolleg 1286 is gratefully acknowledged.


\begin{thebibliography}{33}%
\makeatletter
\providecommand \@ifxundefined [1]{%
 \@ifx{#1\undefined}
}%
\providecommand \@ifnum [1]{%
 \ifnum #1\expandafter \@firstoftwo
 \else \expandafter \@secondoftwo
 \fi
}%
\providecommand \@ifx [1]{%
 \ifx #1\expandafter \@firstoftwo
 \else \expandafter \@secondoftwo
 \fi
}%
\providecommand \natexlab [1]{#1}%
\providecommand \enquote  [1]{``#1''}%
\providecommand \bibnamefont  [1]{#1}%
\providecommand \bibfnamefont [1]{#1}%
\providecommand \citenamefont [1]{#1}%
\providecommand \href@noop [0]{\@secondoftwo}%
\providecommand \href [0]{\begingroup \@sanitize@url \@href}%
\providecommand \@href[1]{\@@startlink{#1}\@@href}%
\providecommand \@@href[1]{\endgroup#1\@@endlink}%
\providecommand \@sanitize@url [0]{\catcode `\\12\catcode `\$12\catcode
  `\&12\catcode `\#12\catcode `\^12\catcode `\_12\catcode `\%12\relax}%
\providecommand \@@startlink[1]{}%
\providecommand \@@endlink[0]{}%
\providecommand \url  [0]{\begingroup\@sanitize@url \@url }%
\providecommand \@url [1]{\endgroup\@href {#1}{\urlprefix }}%
\providecommand \urlprefix  [0]{URL }%
\providecommand \Eprint [0]{\href }%
\providecommand \doibase [0]{http://dx.doi.org/}%
\providecommand \selectlanguage [0]{\@gobble}%
\providecommand \bibinfo  [0]{\@secondoftwo}%
\providecommand \bibfield  [0]{\@secondoftwo}%
\providecommand \translation [1]{[#1]}%
\providecommand \BibitemOpen [0]{}%
\providecommand \bibitemStop [0]{}%
\providecommand \bibitemNoStop [0]{.\EOS\space}%
\providecommand \EOS [0]{\spacefactor3000\relax}%
\providecommand \BibitemShut  [1]{\csname bibitem#1\endcsname}%
\let\auto@bib@innerbib\@empty
\bibitem [{\citenamefont {Kruglyak}\ \emph {et~al.}(2010)\citenamefont
  {Kruglyak}, \citenamefont {Demokritov},\ and\ \citenamefont
  {Grundler}}]{KruglyakJOPD2010}%
  \BibitemOpen
  \bibfield  {author} {\bibinfo {author} {\bibfnamefont {V.~V.}\ \bibnamefont
  {Kruglyak}}, \bibinfo {author} {\bibfnamefont {S.~O.}\ \bibnamefont
  {Demokritov}}, \ and\ \bibinfo {author} {\bibfnamefont {D.}~\bibnamefont
  {Grundler}},\ }\href@noop {} {\bibfield  {journal} {\bibinfo  {journal}
  {Journal of Physics D: Applied Physics}\ }\textbf {\bibinfo {volume} {43}},\
  \bibinfo {pages} {264001} (\bibinfo {year} {2010})}\BibitemShut {NoStop}%
\bibitem [{\citenamefont {Gerdau}\ \emph {et~al.}(1985)\citenamefont {Gerdau},
  \citenamefont {R\"uffer}, \citenamefont {Winkler}, \citenamefont {Tolksdorf},
  \citenamefont {Klages},\ and\ \citenamefont {Hannon}}]{GerdauPRL1985}%
  \BibitemOpen
  \bibfield  {author} {\bibinfo {author} {\bibfnamefont {E.}~\bibnamefont
  {Gerdau}}, \bibinfo {author} {\bibfnamefont {R.}~\bibnamefont {R\"uffer}},
  \bibinfo {author} {\bibfnamefont {H.}~\bibnamefont {Winkler}}, \bibinfo
  {author} {\bibfnamefont {W.}~\bibnamefont {Tolksdorf}}, \bibinfo {author}
  {\bibfnamefont {C.~P.}\ \bibnamefont {Klages}}, \ and\ \bibinfo {author}
  {\bibfnamefont {J.~P.}\ \bibnamefont {Hannon}},\ }\href@noop {} {\bibfield
  {journal} {\bibinfo  {journal} {Phys. Rev. Lett.}\ }\textbf {\bibinfo
  {volume} {54}},\ \bibinfo {pages} {835} (\bibinfo {year} {1985})}\BibitemShut
  {NoStop}%
\bibitem [{\citenamefont {R\"ohlsberger}(2005)}]{RoehlsbergerSpringer2005}%
  \BibitemOpen
  \bibfield  {author} {\bibinfo {author} {\bibfnamefont {R.}~\bibnamefont
  {R\"ohlsberger}},\ }\href@noop {} {\emph {\bibinfo {title} {{Nuclear
  Condensed Matter Physics with Synchrotron Radiation}}}},\ \bibinfo {series}
  {Springer Tracts in Modern Physics}, Vol.\ \bibinfo {volume} {208}\ (\bibinfo
   {publisher} {Springer Berlin Heidelberg},\ \bibinfo {year}
  {2005})\BibitemShut {NoStop}%
\bibitem [{\citenamefont {Schlage}\ \emph {et~al.}(2012)\citenamefont
  {Schlage}, \citenamefont {Couet}, \citenamefont {Roth}, \citenamefont
  {Vainio}, \citenamefont {R\"uffer}, \citenamefont {Kashem}, \citenamefont
  {M\"uller-Buschbaum},\ and\ \citenamefont {R\"ohlsberger}}]{SchlageNJP2012}%
  \BibitemOpen
  \bibfield  {author} {\bibinfo {author} {\bibfnamefont {K.}~\bibnamefont
  {Schlage}}, \bibinfo {author} {\bibfnamefont {S.}~\bibnamefont {Couet}},
  \bibinfo {author} {\bibfnamefont {S.~V.}\ \bibnamefont {Roth}}, \bibinfo
  {author} {\bibfnamefont {U.}~\bibnamefont {Vainio}}, \bibinfo {author}
  {\bibfnamefont {R.}~\bibnamefont {R\"uffer}}, \bibinfo {author}
  {\bibfnamefont {M.~M.~A.}\ \bibnamefont {Kashem}}, \bibinfo {author}
  {\bibfnamefont {P.}~\bibnamefont {M\"uller-Buschbaum}}, \ and\ \bibinfo
  {author} {\bibfnamefont {R.}~\bibnamefont {R\"ohlsberger}},\ }\href@noop {}
  {\bibfield  {journal} {\bibinfo  {journal} {New J. Phys.}\ }\textbf {\bibinfo
  {volume} {14}},\ \bibinfo {pages} {043007} (\bibinfo {year}
  {2012})}\BibitemShut {NoStop}%
\bibitem [{\citenamefont {R\"ohlsberger}\ \emph {et~al.}(2002)\citenamefont
  {R\"ohlsberger}, \citenamefont {Thomas}, \citenamefont {Schlage},
  \citenamefont {Burkel}, \citenamefont {Leupold},\ and\ \citenamefont
  {R\"uffer}}]{RoehlsbergerPRL2002}%
  \BibitemOpen
  \bibfield  {author} {\bibinfo {author} {\bibfnamefont {R.}~\bibnamefont
  {R\"ohlsberger}}, \bibinfo {author} {\bibfnamefont {H.}~\bibnamefont
  {Thomas}}, \bibinfo {author} {\bibfnamefont {K.}~\bibnamefont {Schlage}},
  \bibinfo {author} {\bibfnamefont {E.}~\bibnamefont {Burkel}}, \bibinfo
  {author} {\bibfnamefont {O.}~\bibnamefont {Leupold}}, \ and\ \bibinfo
  {author} {\bibfnamefont {R.}~\bibnamefont {R\"uffer}},\ }\href@noop {}
  {\bibfield  {journal} {\bibinfo  {journal} {Phys. Rev. Lett.}\ }\textbf
  {\bibinfo {volume} {89}},\ \bibinfo {pages} {237201} (\bibinfo {year}
  {2002})}\BibitemShut {NoStop}%
\bibitem [{\citenamefont {Eibsch\"utz}\ and\ \citenamefont
  {Lines}(1973)}]{EibschuetzPRB1973}%
  \BibitemOpen
  \bibfield  {author} {\bibinfo {author} {\bibfnamefont {M.}~\bibnamefont
  {Eibsch\"utz}}\ and\ \bibinfo {author} {\bibfnamefont {M.~E.}\ \bibnamefont
  {Lines}},\ }\href@noop {} {\bibfield  {journal} {\bibinfo  {journal} {Phys.
  Rev. B}\ }\textbf {\bibinfo {volume} {7}},\ \bibinfo {pages} {4907} (\bibinfo
  {year} {1973})}\BibitemShut {NoStop}%
\bibitem [{\citenamefont {Chien}\ and\ \citenamefont
  {Hasegawa}(1977)}]{ChienPRB1977}%
  \BibitemOpen
  \bibfield  {author} {\bibinfo {author} {\bibfnamefont {C.~L.}\ \bibnamefont
  {Chien}}\ and\ \bibinfo {author} {\bibfnamefont {R.}~\bibnamefont
  {Hasegawa}},\ }\href@noop {} {\bibfield  {journal} {\bibinfo  {journal}
  {Phys. Rev. B}\ }\textbf {\bibinfo {volume} {16}},\ \bibinfo {pages} {2115}
  (\bibinfo {year} {1977})}\BibitemShut {NoStop}%
\bibitem [{\citenamefont {Senz}\ \emph {et~al.}(2003)\citenamefont {Senz},
  \citenamefont {R\"ohlsberger}, \citenamefont {Bansmann}, \citenamefont
  {Leopold},\ and\ \citenamefont {Meiwes-Broer}}]{SenzNJP2003}%
  \BibitemOpen
  \bibfield  {author} {\bibinfo {author} {\bibfnamefont {V.}~\bibnamefont
  {Senz}}, \bibinfo {author} {\bibfnamefont {R.}~\bibnamefont {R\"ohlsberger}},
  \bibinfo {author} {\bibfnamefont {J.}~\bibnamefont {Bansmann}}, \bibinfo
  {author} {\bibfnamefont {O.}~\bibnamefont {Leopold}}, \ and\ \bibinfo
  {author} {\bibfnamefont {K.-H.}\ \bibnamefont {Meiwes-Broer}},\ }\href@noop
  {} {\bibfield  {journal} {\bibinfo  {journal} {New J. Phys.}\ }\textbf
  {\bibinfo {volume} {5}},\ \bibinfo {pages} {47} (\bibinfo {year}
  {2003})}\BibitemShut {NoStop}%
\bibitem [{\citenamefont {Demokritov}\ \emph {et~al.}(2006)\citenamefont
  {Demokritov}, \citenamefont {Demidov}, \citenamefont {Dzyapko}, \citenamefont
  {Melkov}, \citenamefont {Serga}, \citenamefont {Hillebrands},\ and\
  \citenamefont {Slavin}}]{DemokritovNAT2006}%
  \BibitemOpen
  \bibfield  {author} {\bibinfo {author} {\bibfnamefont {S.~O.}\ \bibnamefont
  {Demokritov}}, \bibinfo {author} {\bibfnamefont {V.~E.}\ \bibnamefont
  {Demidov}}, \bibinfo {author} {\bibfnamefont {O.}~\bibnamefont {Dzyapko}},
  \bibinfo {author} {\bibfnamefont {G.~A.}\ \bibnamefont {Melkov}}, \bibinfo
  {author} {\bibfnamefont {A.~A.}\ \bibnamefont {Serga}}, \bibinfo {author}
  {\bibfnamefont {B.}~\bibnamefont {Hillebrands}}, \ and\ \bibinfo {author}
  {\bibfnamefont {A.~N.}\ \bibnamefont {Slavin}},\ }\href@noop {} {\bibfield
  {journal} {\bibinfo  {journal} {Nature}\ }\textbf {\bibinfo {volume} {443}},\
  \bibinfo {pages} {430} (\bibinfo {year} {2006})}\BibitemShut {NoStop}%
\bibitem [{\citenamefont {Kostylev}\ \emph {et~al.}(2005)\citenamefont
  {Kostylev}, \citenamefont {Serga}, \citenamefont {Schneider}, \citenamefont
  {Leven},\ and\ \citenamefont {Hillebrands}}]{KostylevAPL2005}%
  \BibitemOpen
  \bibfield  {author} {\bibinfo {author} {\bibfnamefont {M.~P.}\ \bibnamefont
  {Kostylev}}, \bibinfo {author} {\bibfnamefont {A.~A.}\ \bibnamefont {Serga}},
  \bibinfo {author} {\bibfnamefont {T.}~\bibnamefont {Schneider}}, \bibinfo
  {author} {\bibfnamefont {B.}~\bibnamefont {Leven}}, \ and\ \bibinfo {author}
  {\bibfnamefont {B.}~\bibnamefont {Hillebrands}},\ }\href@noop {} {\bibfield
  {journal} {\bibinfo  {journal} {Appl. Phys. Lett.}\ }\textbf {\bibinfo
  {volume} {87}},\ \bibinfo {pages} {153501} (\bibinfo {year}
  {2005})}\BibitemShut {NoStop}%
\bibitem [{\citenamefont {Sinor}\ \emph {et~al.}(1989)\citenamefont {Sinor},
  \citenamefont {Reittinger},\ and\ \citenamefont {Collins}}]{SinorPRL1989}%
  \BibitemOpen
  \bibfield  {author} {\bibinfo {author} {\bibfnamefont {T.~W.}\ \bibnamefont
  {Sinor}}, \bibinfo {author} {\bibfnamefont {P.~W.}\ \bibnamefont
  {Reittinger}}, \ and\ \bibinfo {author} {\bibfnamefont {C.~B.}\ \bibnamefont
  {Collins}},\ }\href@noop {} {\bibfield  {journal} {\bibinfo  {journal} {Phys.
  Rev. Lett.}\ }\textbf {\bibinfo {volume} {62}},\ \bibinfo {pages} {2547}
  (\bibinfo {year} {1989})}\BibitemShut {NoStop}%
\bibitem [{\citenamefont {Pfeiffer}(1971)}]{PfeifferJAP1971}%
  \BibitemOpen
  \bibfield  {author} {\bibinfo {author} {\bibfnamefont {L.}~\bibnamefont
  {Pfeiffer}},\ }\href@noop {} {\bibfield  {journal} {\bibinfo  {journal}
  {Phys. Rev. B}\ }\textbf {\bibinfo {volume} {42}},\ \bibinfo {pages} {1725}
  (\bibinfo {year} {1971})}\BibitemShut {NoStop}%
\bibitem [{\citenamefont {Kopcewicz}(1991)}]{KopcewiczSC1991}%
  \BibitemOpen
  \bibfield  {author} {\bibinfo {author} {\bibfnamefont {M.}~\bibnamefont
  {Kopcewicz}},\ }\href@noop {} {\bibfield  {journal} {\bibinfo  {journal}
  {Struct. Chem.}\ }\textbf {\bibinfo {volume} {2}},\ \bibinfo {pages} {313}
  (\bibinfo {year} {1991})}\BibitemShut {NoStop}%
\bibitem [{\citenamefont {Heiman}\ \emph {et~al.}(1968)\citenamefont {Heiman},
  \citenamefont {Pfeiffer},\ and\ \citenamefont {Walker}}]{HeimanPRL1968}%
  \BibitemOpen
  \bibfield  {author} {\bibinfo {author} {\bibfnamefont {N.~D.}\ \bibnamefont
  {Heiman}}, \bibinfo {author} {\bibfnamefont {L.}~\bibnamefont {Pfeiffer}}, \
  and\ \bibinfo {author} {\bibfnamefont {J.~C.}\ \bibnamefont {Walker}},\
  }\href@noop {} {\bibfield  {journal} {\bibinfo  {journal} {Phys. Rev. Lett.}\
  }\textbf {\bibinfo {volume} {21}},\ \bibinfo {pages} {93} (\bibinfo {year}
  {1968})}\BibitemShut {NoStop}%
\bibitem [{\citenamefont {Tittonen}\ \emph {et~al.}(1992)\citenamefont
  {Tittonen}, \citenamefont {Lippmaa}, \citenamefont {Ikonen}, \citenamefont
  {Lind\'en},\ and\ \citenamefont {Katila}}]{TittonenPRL1992}%
  \BibitemOpen
  \bibfield  {author} {\bibinfo {author} {\bibfnamefont {I.}~\bibnamefont
  {Tittonen}}, \bibinfo {author} {\bibfnamefont {M.}~\bibnamefont {Lippmaa}},
  \bibinfo {author} {\bibfnamefont {E.}~\bibnamefont {Ikonen}}, \bibinfo
  {author} {\bibfnamefont {J.}~\bibnamefont {Lind\'en}}, \ and\ \bibinfo
  {author} {\bibfnamefont {T.}~\bibnamefont {Katila}},\ }\href@noop {}
  {\bibfield  {journal} {\bibinfo  {journal} {Phys. Rev. Lett.}\ }\textbf
  {\bibinfo {volume} {69}},\ \bibinfo {pages} {2815} (\bibinfo {year}
  {1992})}\BibitemShut {NoStop}%
\bibitem [{\citenamefont {Tserkovnyak}\ \emph {et~al.}(2005)\citenamefont
  {Tserkovnyak}, \citenamefont {Brataas}, \citenamefont {Bauer},\ and\
  \citenamefont {Halperin}}]{TserkovnyakPMP2005}%
  \BibitemOpen
  \bibfield  {author} {\bibinfo {author} {\bibfnamefont {Y.}~\bibnamefont
  {Tserkovnyak}}, \bibinfo {author} {\bibfnamefont {A.}~\bibnamefont
  {Brataas}}, \bibinfo {author} {\bibfnamefont {G.~E.~W.}\ \bibnamefont
  {Bauer}}, \ and\ \bibinfo {author} {\bibfnamefont {B.~I.}\ \bibnamefont
  {Halperin}},\ }\href@noop {} {\bibfield  {journal} {\bibinfo  {journal} {Rev.
  Mod. Phys.}\ }\textbf {\bibinfo {volume} {77}},\ \bibinfo {pages} {1375}
  (\bibinfo {year} {2005})}\BibitemShut {NoStop}%
\bibitem [{\citenamefont {Kuhlmann}\ \emph {et~al.}(2013)\citenamefont
  {Kuhlmann}, \citenamefont {Swoboda}, \citenamefont {Vogel}, \citenamefont
  {Matsuyama},\ and\ \citenamefont {Meier}}]{KuhlmannPRB2013}%
  \BibitemOpen
  \bibfield  {author} {\bibinfo {author} {\bibfnamefont {N.}~\bibnamefont
  {Kuhlmann}}, \bibinfo {author} {\bibfnamefont {C.}~\bibnamefont {Swoboda}},
  \bibinfo {author} {\bibfnamefont {A.}~\bibnamefont {Vogel}}, \bibinfo
  {author} {\bibfnamefont {T.}~\bibnamefont {Matsuyama}}, \ and\ \bibinfo
  {author} {\bibfnamefont {G.}~\bibnamefont {Meier}},\ }\href@noop {}
  {\bibfield  {journal} {\bibinfo  {journal} {Phys. Rev. B}\ }\textbf {\bibinfo
  {volume} {87}},\ \bibinfo {pages} {104409} (\bibinfo {year}
  {2013})}\BibitemShut {NoStop}%
\bibitem [{\citenamefont {Gui}\ \emph {et~al.}(2007)\citenamefont {Gui},
  \citenamefont {Mecking}, \citenamefont {Zhou}, \citenamefont {Williams},\
  and\ \citenamefont {Hu}}]{GuiPRL2007}%
  \BibitemOpen
  \bibfield  {author} {\bibinfo {author} {\bibfnamefont {Y.~S.}\ \bibnamefont
  {Gui}}, \bibinfo {author} {\bibfnamefont {N.}~\bibnamefont {Mecking}},
  \bibinfo {author} {\bibfnamefont {X.}~\bibnamefont {Zhou}}, \bibinfo {author}
  {\bibfnamefont {G.}~\bibnamefont {Williams}}, \ and\ \bibinfo {author}
  {\bibfnamefont {C.-M.}\ \bibnamefont {Hu}},\ }\href@noop {} {\bibfield
  {journal} {\bibinfo  {journal} {Phys. Rev. Lett.}\ }\textbf {\bibinfo
  {volume} {98}},\ \bibinfo {pages} {107602} (\bibinfo {year}
  {2007})}\BibitemShut {NoStop}%
\bibitem [{\citenamefont {Costache}\ \emph {et~al.}(2006)\citenamefont
  {Costache}, \citenamefont {Watts}, \citenamefont {Sladkov}, \citenamefont
  {van~der Wal},\ and\ \citenamefont {van Wees}}]{CostacheAPL06b}%
  \BibitemOpen
  \bibfield  {author} {\bibinfo {author} {\bibfnamefont {M.~V.}\ \bibnamefont
  {Costache}}, \bibinfo {author} {\bibfnamefont {S.~M.}\ \bibnamefont {Watts}},
  \bibinfo {author} {\bibfnamefont {M.}~\bibnamefont {Sladkov}}, \bibinfo
  {author} {\bibfnamefont {C.~H.}\ \bibnamefont {van~der Wal}}, \ and\ \bibinfo
  {author} {\bibfnamefont {B.~J.}\ \bibnamefont {van Wees}},\ }\href@noop {}
  {\bibfield  {journal} {\bibinfo  {journal} {Appl. Phys. Lett.}\ }\textbf
  {\bibinfo {volume} {89}},\ \bibinfo {pages} {232115} (\bibinfo {year}
  {2006})}\BibitemShut {NoStop}%
\bibitem [{\citenamefont {Kuhlmann}\ \emph {et~al.}(2012)\citenamefont
  {Kuhlmann}, \citenamefont {Vogel},\ and\ \citenamefont
  {Meier}}]{KuhlmannPRB2012}%
  \BibitemOpen
  \bibfield  {author} {\bibinfo {author} {\bibfnamefont {N.}~\bibnamefont
  {Kuhlmann}}, \bibinfo {author} {\bibfnamefont {A.}~\bibnamefont {Vogel}}, \
  and\ \bibinfo {author} {\bibfnamefont {G.}~\bibnamefont {Meier}},\
  }\href@noop {} {\bibfield  {journal} {\bibinfo  {journal} {Phys. Rev. B}\
  }\textbf {\bibinfo {volume} {85}},\ \bibinfo {pages} {014410} (\bibinfo
  {year} {2012})}\BibitemShut {NoStop}%
\bibitem [{\citenamefont {Blume}(1968)}]{BlumePR1968b}%
  \BibitemOpen
  \bibfield  {author} {\bibinfo {author} {\bibfnamefont {M.}~\bibnamefont
  {Blume}},\ }\href@noop {} {\bibfield  {journal} {\bibinfo  {journal} {Phys.
  Rev.}\ }\textbf {\bibinfo {volume} {174}},\ \bibinfo {pages} {351} (\bibinfo
  {year} {1968})}\BibitemShut {NoStop}%
\bibitem [{\citenamefont {Clauser}\ and\ \citenamefont
  {Blume}(1971)}]{ClauserPRB1971}%
  \BibitemOpen
  \bibfield  {author} {\bibinfo {author} {\bibfnamefont {M.~J.}\ \bibnamefont
  {Clauser}}\ and\ \bibinfo {author} {\bibfnamefont {M.}~\bibnamefont
  {Blume}},\ }\href@noop {} {\bibfield  {journal} {\bibinfo  {journal} {Phys.
  Rev. B}\ }\textbf {\bibinfo {volume} {3}},\ \bibinfo {pages} {583} (\bibinfo
  {year} {1971})}\BibitemShut {NoStop}%
\bibitem [{\citenamefont {Smit}\ and\ \citenamefont
  {Beljers}(1955)}]{SmitPRR1955}%
  \BibitemOpen
  \bibfield  {author} {\bibinfo {author} {\bibfnamefont {J.}~\bibnamefont
  {Smit}}\ and\ \bibinfo {author} {\bibfnamefont {H.~G.}\ \bibnamefont
  {Beljers}},\ }\href@noop {} {\bibfield  {journal} {\bibinfo  {journal}
  {Philips Research Reports}\ }\textbf {\bibinfo {volume} {10}},\ \bibinfo
  {pages} {113} (\bibinfo {year} {1955})}\BibitemShut {NoStop}%
\bibitem [{\citenamefont {Wille}\ \emph {et~al.}(2010)\citenamefont {Wille},
  \citenamefont {Franz}, \citenamefont {R\"ohlsberger}, \citenamefont
  {Caliebe},\ and\ \citenamefont {Dill}}]{WilleJPCS2010}%
  \BibitemOpen
  \bibfield  {author} {\bibinfo {author} {\bibfnamefont {H.-C.}\ \bibnamefont
  {Wille}}, \bibinfo {author} {\bibfnamefont {H.}~\bibnamefont {Franz}},
  \bibinfo {author} {\bibfnamefont {R.}~\bibnamefont {R\"ohlsberger}}, \bibinfo
  {author} {\bibfnamefont {W.~A.}\ \bibnamefont {Caliebe}}, \ and\ \bibinfo
  {author} {\bibfnamefont {F.-U.}\ \bibnamefont {Dill}},\ }\href@noop {}
  {\bibfield  {journal} {\bibinfo  {journal} {Journal of Physics: Conference
  Series}\ }\textbf {\bibinfo {volume} {217}},\ \bibinfo {pages} {012008}
  (\bibinfo {year} {2010})},\ \bibinfo {note}
  {http://photon-science.desy.de/\-facilities/\-petra\_iii/\-beamlines/\-p01\_dynamics/\-index\_eng.html}\BibitemShut
  {NoStop}%
\bibitem [{Sup()}]{Supplementary}%
  \BibitemOpen
  \href@noop {} {}\bibinfo {note} {For more infoprmation see suplementary
  material.}\BibitemShut {Stop}%
\bibitem [{\citenamefont {Kittel}(1949)}]{KittelRMP49}%
  \BibitemOpen
  \bibfield  {author} {\bibinfo {author} {\bibfnamefont {C.}~\bibnamefont
  {Kittel}},\ }\href@noop {} {\bibfield  {journal} {\bibinfo  {journal} {Rev.
  Mod. Phys.}\ }\textbf {\bibinfo {volume} {21}},\ \bibinfo {pages} {541}
  (\bibinfo {year} {1949})}\BibitemShut {NoStop}%
\bibitem [{\citenamefont {Kamionka}\ \emph {et~al.}(2011)\citenamefont
  {Kamionka}, \citenamefont {Martens}, \citenamefont {Drews}, \citenamefont
  {Kr\"uger}, \citenamefont {Albrecht},\ and\ \citenamefont
  {Meier}}]{KamionkaPRB2011}%
  \BibitemOpen
  \bibfield  {author} {\bibinfo {author} {\bibfnamefont {T.}~\bibnamefont
  {Kamionka}}, \bibinfo {author} {\bibfnamefont {M.}~\bibnamefont {Martens}},
  \bibinfo {author} {\bibfnamefont {A.}~\bibnamefont {Drews}}, \bibinfo
  {author} {\bibfnamefont {B.}~\bibnamefont {Kr\"uger}}, \bibinfo {author}
  {\bibfnamefont {O.}~\bibnamefont {Albrecht}}, \ and\ \bibinfo {author}
  {\bibfnamefont {G.}~\bibnamefont {Meier}},\ }\href@noop {} {\bibfield
  {journal} {\bibinfo  {journal} {Phys. Rev. B}\ }\textbf {\bibinfo {volume}
  {83}},\ \bibinfo {pages} {224424} (\bibinfo {year} {2011})}\BibitemShut
  {NoStop}%
\bibitem [{\citenamefont {Stuhrhahn}(2000)}]{SturhahnHI2000}%
  \BibitemOpen
  \bibfield  {author} {\bibinfo {author} {\bibfnamefont {W.}~\bibnamefont
  {Stuhrhahn}},\ }\href@noop {} {\bibfield  {journal} {\bibinfo  {journal}
  {Hyperfine Interact.}\ }\textbf {\bibinfo {volume} {125}},\ \bibinfo {pages}
  {237201} (\bibinfo {year} {2000})}\BibitemShut {NoStop}%
\bibitem [{\citenamefont {Ping}\ \emph {et~al.}(1992)\citenamefont {Ping},
  \citenamefont {Rancourt},\ and\ \citenamefont {Dunlap}}]{PingJMMM1992}%
  \BibitemOpen
  \bibfield  {author} {\bibinfo {author} {\bibfnamefont {J.}~\bibnamefont
  {Ping}}, \bibinfo {author} {\bibfnamefont {D.}~\bibnamefont {Rancourt}}, \
  and\ \bibinfo {author} {\bibfnamefont {R.}~\bibnamefont {Dunlap}},\
  }\href@noop {} {\bibfield  {journal} {\bibinfo  {journal} {J. Magn. Magn.
  Mat.}\ }\textbf {\bibinfo {volume} {103}},\ \bibinfo {pages} {285 } (\bibinfo
  {year} {1992})}\BibitemShut {NoStop}%
\bibitem [{\citenamefont {Blume}\ and\ \citenamefont
  {Tjon}(1968)}]{BlumePR1968}%
  \BibitemOpen
  \bibfield  {author} {\bibinfo {author} {\bibfnamefont {M.}~\bibnamefont
  {Blume}}\ and\ \bibinfo {author} {\bibfnamefont {J.~A.}\ \bibnamefont
  {Tjon}},\ }\href@noop {} {\bibfield  {journal} {\bibinfo  {journal} {Phys.
  Rev.}\ }\textbf {\bibinfo {volume} {165}},\ \bibinfo {pages} {446} (\bibinfo
  {year} {1968})}\BibitemShut {NoStop}%
\bibitem [{\citenamefont {Leupold}\ and\ \citenamefont
  {Winkler}(1999)}]{LeopoldHI1999}%
  \BibitemOpen
  \bibfield  {author} {\bibinfo {author} {\bibfnamefont {O.}~\bibnamefont
  {Leupold}}\ and\ \bibinfo {author} {\bibfnamefont {H.}~\bibnamefont
  {Winkler}},\ }\href@noop {} {\bibfield  {journal} {\bibinfo  {journal}
  {Hyperfine Interactions}\ }\textbf {\bibinfo {volume} {123-124}},\ \bibinfo
  {pages} {571} (\bibinfo {year} {1999})}\BibitemShut {NoStop}%
\bibitem [{\citenamefont {Skrotskii}\ and\ \citenamefont
  {Kurbatov}(1966)}]{SkrotskiiFR1966}%
  \BibitemOpen
  \bibfield  {author} {\bibinfo {author} {\bibfnamefont {G.}~\bibnamefont
  {Skrotskii}}\ and\ \bibinfo {author} {\bibfnamefont {L.}~\bibnamefont
  {Kurbatov}},\ }in\ \href {\doibase
  http://dx.doi.org/10.1016/B978-0-08-011027-1.50005-7} {\emph {\bibinfo
  {booktitle} {Ferromagnetic Resonance}}},\ \bibinfo {editor} {edited by\
  \bibinfo {editor} {\bibfnamefont {S.}~\bibnamefont {Vonsovskii}}}\ (\bibinfo
  {publisher} {Pergamon},\ \bibinfo {year} {1966})\ pp.\ \bibinfo {pages} {12
  -- 77}\BibitemShut {NoStop}%
\bibitem [{Mic()}]{MicroMagnum}%
  \BibitemOpen
  \href@noop {} {}\bibinfo {note} {MicroMagnum Fast Physical Simulator for
  Computations on CPU and Graphics Processing Unit
  http://micromagnum.informatik.uni-hamburg.de/.}\BibitemShut {Stop}%
\end{thebibliography}
\end{document}